\documentstyle[preprint,aps,epsf]{revtex}
\begin{document}
\tightenlines
\draft
\title{Bethe Approximation for Self-Interacting Lattice Trees}
\author{
Paolo De Los Rios$^{1,a}$,
Stefano Lise$^{2,b}$ and
Alessandro Pelizzola$^{3,c}$ }
\address{(1) Institut de Physique Theorique, Universit\'e de Fribourg, 
Chemin du Mus\'ee 3, \\ CH-1700 Fribourg, Switzerland,    }
\address{(2) Department of Mathematics, Imperial College, 
 180 Queen's Gate, London SW7 2BZ, UK  }
\address{(3) Dipartimento di Fisica, Politecnico di Torino,
c. Duca degli Abruzzi 24, 10129 Torino, Italy  \\
and Istituto Nazionale per la Fisica della Materia, Unit\`a Torino
Politecnico.} 
\date{\today}
\maketitle
\begin{abstract}
In this paper we develop a Bethe approximation, based on the cluster 
variation method, which is apt to study lattice models of branched 
polymers. We show that the method is extremely accurate in cases where
exact results are known as, for instance, in the enumeration of spanning 
trees. Moreover, the expressions we obtain for the asymptotic number of
spanning trees and lattice trees on a graph coincide with analogous
expressions derived through different approaches.
We study the phase diagram of lattice trees with nearest-neighbour 
attraction and branching energies.  We find a collapse transition at a
tricritical $\theta$ point, 
which separates an expanded phase from a compact phase. We compare our
results for the $\theta$ transition in two and three dimensions with
available numerical estimates.
\vskip 0.3cm
\noindent PACS numbers: 5.70.Fh, 36.20.Ey, 64.60.Cn
\vskip 0.6cm
\end{abstract}
\narrowtext

The statistical properties of polymers have been one of the great
challenges of statistical physics in the last
decades\cite{deg_1,des}. A great understanding has been gained
for the case of linear polymers, thanks to the simple but instructive
description in terms of self-avoiding walks (SAWs) on a 
lattice, which is in turn amenable to many different 
treatments, ranging from Monte Carlo simulations to exact
enumerations, from mapping to the $O(n=0)$ spin model to field 
theoretical formulations\cite{vander}.  
Less is known about the behaviour of branched polymers (BPs). In 
lattice statistical mechanics BPs can be efficiently modelled by lattice 
animals\cite{vander}, i.e. by connected clusters of bonds. 
Analogously to linear polymers, BPs in solution can display a dense and a 
diluted phase, depending on temperature and quality of the solvent.
The collapse transition occurs at the so called $\theta$ 
point\cite{deg_2}  and in recent years has attracted a lot of 
attention, both for linear \cite{seno_1}  and branched polymers\cite{vander}.

Recently, a variational approach based on the cluster variation method 
(CVM) has been introduced for linear polymers\cite{lise}, giving results 
in good agreement with the best numerical simulations in many situations: 
dense polymers (i.e. Hamiltonian Walks), diluted and self-interacting 
ones. The CVM \cite{kik1,an} is a closed form approximation, which finds 
wide applications in accurate investigations  of the phase diagram of 
lattice spin systems \cite{cvm}. It is based on the minimisation of a
variational free energy which is obtained by truncating the
lattice cumulant expansion of the entropy \cite{an}. The largest 
clusters considered in the expansion determine the ``order'' of the 
approximation and are named maximal clusters. For example, standard 
mean-field and Bethe approximations are recovered by considering as 
maximal clusters respectively single sites and nearest-neighbours pairs.
Most of the present knowledge about the phase behaviour of BPs has been 
gained through numerical approaches\cite{vander} such as Monte Carlo 
simulations, exact enumerations or transfer matrix techniques (for 
an exception see, e.g., ref.~\cite{henkel}).
The possibility to extend to BPs the CVM scheme  becomes therefore 
extremely appealing.

In the present paper we introduce a CVM based variational approximation
for lattice trees (LTs), that is lattice animals containing no closed 
loops. 
We show that the method performs very well when compared to situations
where exact results are known as, for instance, in the case of the 
enumeration of spanning trees on a lattice. 
We then use it to explore the phase behaviour of LTs with branching
energies and nearest-neighbour
attractive interaction between non bonded monomers, in different energy 
and temperature regimes.

Let us make first a rather important technical remark about our method. 
Being self-avoiding, LTs face strong non-local constraints on the lattice
(neighbouring sites can be occupied by monomers that are
extremely far apart on the tree). This is a serious
problem when dealing with local approximations such as the CVM.
In order to partially overcome this difficulty,
we have introduced an intrinsic direction on  
each edge of a lattice tree, which can be visualised with an arrow. 
Each occupied site of the lattice is therefore characterised by a certain 
number of incoming edges (input edges) and by a certain number of
outgoing edges (output edges). To ensure that the tree does not form loops 
we have required that (i) sites with no outputs occur with probability
0 in the thermodynamic limit and 
(ii) all other occupied sites have exactly one and only one output. 
It is important to underline that conditions (i) and (ii) uniquely determine 
the orientation of the arrows on a given LT, so that the descriptions in
terms of oriented LTs and non oriented LTs are equivalent. The advantage for
our purposes is that we are able in this way to exclude configurations
with an inner loop
(which is a kind of long range constraint) by means of a local 
variable (the orientation of the arrow on each occupied bond). 

Within the CVM scheme, the Bethe approximation is obtained by choosing
nearest neighbour (NN) pairs as maximal clusters. For a homogeneous and 
isotropic model with only single site and NN pair interaction terms, the
variational (reduced) free energy density has the form \cite{an}
\begin{equation}
\label{free_en}
\tilde f ^{\mbox{\tiny $(B)$}} = 
\frac{\beta F^{\mbox{\tiny $(B)$}} }{N} = 
\frac{q}{2} \beta {\rm Tr}(\rho_{\rm NN}
{\cal H}_{\rm NN}) + \frac{q}{2} {\rm Tr}(\rho_{\rm NN} \ln \rho_{\rm
NN}) - (q-1) {\rm Tr}(\rho_\bullet \ln \rho_\bullet),
\end{equation}
where $q$ is the coordination number of the lattice, $\beta = 1/k_B
T$ ($k_B = 1$ from now on), ${\cal H}_{\rm NN}$ is the contribution of
a generic NN pair to 
the Hamiltonian and $\rho_\bullet$ and $\rho_{\rm
NN}$ are the site and NN pair density matrices, respectively. For
classical models the density matrices are diagonal and their diagonal
elements are the probability of the corresponding configurations. The
free energy must be minimised with respect to the density matrices,
which must satisfy the condition of normalisation to 1 and
compatibility (that is, $\rho_\bullet$ must be obtainable from
$\rho_{\rm NN}$ by a partial trace).

The first problem we deal with are spanning trees (STs).
A ST visits all sites of a lattice and  therefore STs can be seen as 
a special case (a subset) of LTs. 
In particular, STs are believed to model compact branched polymers. 
For the moment we are interested in estimating the total number ${\cal
N}_{ST}$ of STs 
on a lattice. To leading order, this number should scale with the number 
of sites $N$ as 
$\mbox{$\cal{N}$}_{\mbox{\tiny $ST$}}\sim \mu_{\mbox{\tiny $ST$}}^N$,
where $\mu_{\mbox{\tiny $ST$}}$ is the so called connective constant.
As the entropy per site $s_{\mbox{\tiny $ST$}}$ is given by
$s_{\mbox{\tiny $ST$}} = \ln \mu_{\mbox{\tiny $ST$}}$ and there is no
energy term, it follows that the reduced free energy per
site $\tilde f_{\mbox{\tiny $ST$}} $ is simply  related to the 
connective constant by 
$\tilde f_{\mbox{\tiny $ST$}} = - \ln \mu_{\mbox{\tiny $ST$}}$.
The value of $\mu_{\mbox{\tiny $ST$}}$ for a given graph
is a non trivial number, which can anyway be  calculated exactly. It is 
therefore an excellent starting point to test the accuracy of our method.

\begin{figure}[hb]
\protect\vspace{0.3cm}
 \epsfxsize=3.3in
\centerline{\epsffile{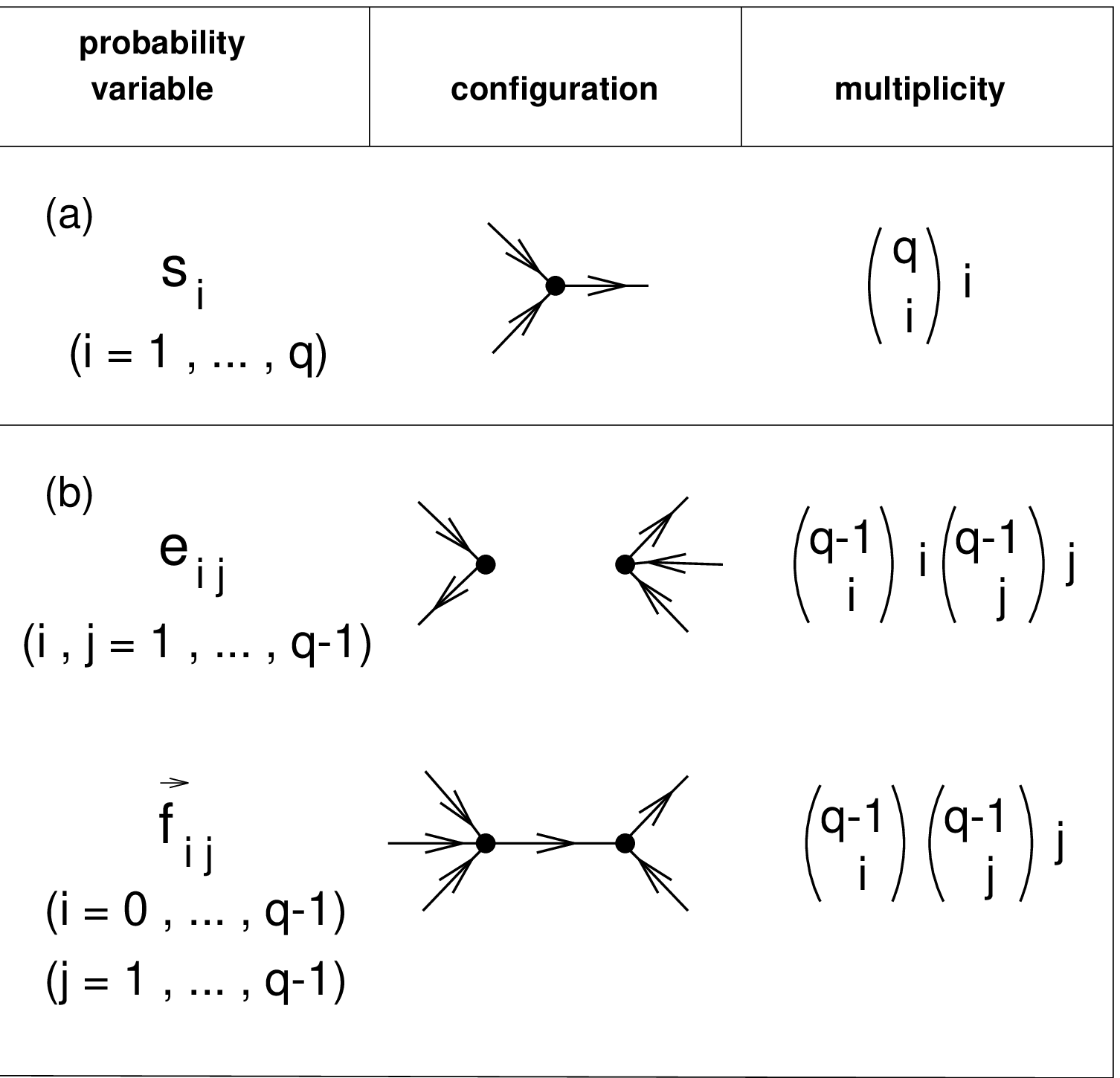}}
\protect\vspace{0.3cm}
\caption[1]{\label{conf_fig}
Schematic representation of independent (a) site
and (b) pair configurations, in the case of spanning
trees. The continuous line represents the spanning tree; 
$q=2d$ is the coordination number of the lattice. 
The configurations drawn in the picture 
correspond to (a) $s_3$, (b) $e_{2,3}$ and $\vec{f}_{3,2}$.  }
\end{figure}

The configurations of a site and a NN pair are classified according
to the number of edges attached to each site and they are reported 
schematically in fig.~\ref{conf_fig}. In the following we will denote with 
$s_i$ the probability variable of a site configuration with $i$ occupied
edges (e.g. for a square lattice $i$ takes values between $1$ and $4$) and 
with $e_{ij}$ the  probability variable assigned to a disconnected pair 
having $i$ edges on one site and $j$ edges on the other site.  
Similarly we will denote with $\vec{f}_{ij}$ the probability variable of 
a connected pair with an intrinsic direction from the site with $i$ edges 
to the site with  $j$ edges.
The free energy (\ref{free_en}) takes then the form
\begin{equation}
\label{free_st1} 
\tilde{f}_{\mbox{\tiny $ST$}}^{\mbox{\tiny $(B)$}} = 
   \frac{q}{2}\left\{ \sum _{i,j=1}^{q-1} m_e(i,j) e_{ij} \ln e_{ij}+
         2 \sum_{i=0}^{q-1} \sum_{j=1}^{q-1} m_f(i,j) 
            \vec{f}_{ij} \ln \vec{f}_{ij} 
             \right\} - 
           (q-1) \left\{ \sum_{i=1}^{q} m_s(i) s_i \ln s_i \right\}   
\end{equation}
where $m_s(i)$, $m_e(i,j)$ and $m_f(i,j)$ stand respectively for the  
multiplicity of site, disconnected pair and connected pair configurations 
(see fig.~\ref{conf_fig} for the actual values of $m_s(i)$, $m_e(i,j)$ and 
$m_f(i,j)$).
The factor $2$ in front of the contribution from the connected        
pairs configurations arises from the degeneracy associated with the 
direction  of the arrow.  
In writing equation (\ref{free_st1}), moreover, we have implicitly assumed 
translational invariance in the system, a condition which in the 
thermodynamic limit is fulfilled.
The normalisation and compatibility conditions on the density matrices
can be written respectively as
\begin {equation}
\sum_{i,j=0}^{q-1} m_e(i,j) e_{ij} + 2 \sum_{i=0}^{q-1}
\sum_{j=1}^{q-1} m_f(i,j) \vec{f}_{ij} = 1 \nonumber 
\end{equation}
and
\begin{eqnarray}
\label{compatib}
\sum_{j=1}^{q-1} {q-1 \choose j} j e_{ij} &=& s_i, 
\qquad i = 1, \ldots q-1 \nonumber \\
\sum_{j=1}^{q-1} {q-1 \choose j} j \vec{f}_{i-1,j} &=& s_i,
\qquad i = 1, \ldots q  \\
\sum_{j=0}^{q-1} {q-1 \choose j} \vec{f}_{j,i-1} &=& s_i
, \qquad i = 2, \ldots q. \nonumber
\end{eqnarray}
Conditions (\ref{compatib}) assures that single site and pair probability
variables are defined consistently, so that by summing over all allowed
configurations of one site of a pair, one should recover the probability
$s_i$ associated to a single site configuration.
Our problem is then to find the minimum of the CVM free energy $\tilde
f_{\mbox{\tiny $ST$}}^{\mbox{\tiny $(B)$}}$ subject to the above constraints. 
In general this problem can be easily dealt with numerically with the help 
of an algorithm named  numerical iteration method \cite{kik2}. In this 
special case of spanning trees enumeration however one can first guess on the
basis of numerical results and then verify by direct 
substitution that the solution takes the analytical form
$s_i =\displaystyle\frac{(q-2)^{q-i}}{q(q-1)^{q-1}}$, $e_{ij} =
\displaystyle\frac{q}{q-2} s_i s_j$, $\vec{f}_{ij} = s_{i+1} s_{j+1}$,
which gives for the reduced free energy
\begin{equation}
\label{free_st2}
\tilde{f}_{\mbox{\tiny $ST$}}^{\mbox{\tiny $(B)$}} = 
\frac{ (q-2) \ln \left[ q(q-2) \right] -2(q-1) \ln (q-1) }{2} 
\label{q site entropy}
\end{equation}
from which one can derive the connective constant 
$ \mu_{\mbox{\tiny $ST$}}^{\mbox{\tiny $(B)$}} = 
\exp (-\tilde{f}_{\mbox{\tiny $ST$}}^{\mbox{\tiny $(B)$}})$.

It is interesting to remark that the same result for the  reduced free 
energy (\ref{free_st2}) can be obtained by exploiting the well-known
relation between STs and the  the $Q$-state Potts model \cite{wu}. 
Indeed by applying the same approximation scheme to the Potts model and then 
taking the $Q \to 0$ limit of the reduced free energy density  calculated in 
$\beta = \ln(1 + Q^\alpha)$, where $\beta$ denotes the Potts coupling 
divided by $k_B T$ and $0 < \alpha < 1$, one recovers expression
(\ref{free_st2}).

Actually, the exact number of STs on a $d$-dimensional hypercubic lattice 
can be computed exactly \cite{harary}. 
The entropy is the logarithm of this number and, in the thermodynamic
limit $N \to \infty$, the entropy per site is
\begin{equation}
s_{\mbox{\tiny $ST$}}= 
\frac{1}{(2 \pi)^d} \int_0^{2 \pi} dk_1 ... \int_0^{2 \pi} dk_d
\ln \left(2d - 2 \sum_{i=1}^d \cos(k_i) \right)
\label{site entropy}
\end{equation}
The large $d$ expansion of (\ref{site entropy})
is  
\begin{equation}
s_{\mbox{\tiny $ST$}} = \ln (2d) - \frac{1}{4d} -\frac{3}{8 d^2}+
 o\left(\frac{1}{d^3}\right)
\label{exact expansion}
\end{equation}
Expanding (\ref{q site entropy}) for large $q$ 
we get
\begin{equation}
s_{\mbox{\tiny $ST$}}^{\mbox{\tiny $(B)$}} =
 \ln q - \frac{1}{2q} - \frac{1}{2 q^2} + o\left(\frac{1}{q^3}\right)
\label{our expansion}
\end{equation}
With $q=2d$ the first three terms (the logarithm, the 
vanishing constant and the first power of $1/d$) coincide.

\begin{figure}[hb]
\epsfxsize=3.7in
\centerline{\epsffile{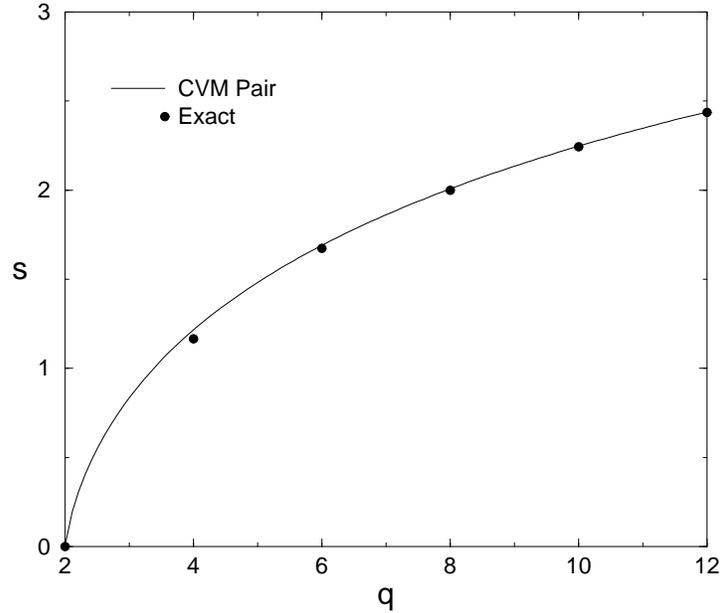}}
\caption[3]{\label{entropy_fig}
Entropy per site as a function of coordination number $q$ in the
case of spanning trees. The continuous line corresponds to
the Bethe approximation, the circles are exact results.}
\end{figure}

In fig.~\ref{entropy_fig} we show a comparison between  
$ s_{\mbox{\tiny $ST$}}^{\mbox{\tiny $(B)$}}$ and       
the exact result $ s_{\mbox{\tiny $ST$}}$, which indicates that
the accuracy of the approximation is rather good.
This check is of utmost importance. Indeed, the CVM is an approximation
on the entropy estimate of the system. Having an extremely good
approximation of the entropy gives confidence in more complicated
situations where energies are introduced into the model.
The probability $p_i$ that a randomly chosen site is connected 
to $i$ nearest-neighbour sites can also be calculated exactly for STs on 
the square lattice\cite{manna}. The result is 
$p_1 \approx 0.29454$, $p_2 \approx 0.44699$, $p_3 \approx 0.22239$ and 
$p_4 \approx 0.03608$, which agree quite well with the values
we obtain within our approximation ($p_i=m_s(i) s_i$), i.e.  
$p_1 \approx 0.29630$, $p_2 \approx 0.44444$, $p_3 \approx 0.22222$ and 
$p_4 \approx 0.03704$.

As a further test, we have considered the case of directed  spanning trees 
(DSTs) in $d$ dimensions, where the exact result is known and it is simple. 
In this case a preferred direction along a lattice diagonal is chosen and 
all edges must have a positive component along this direction. This means 
that the output at each site is restricted to $d$ possible directions. 
Moreover the output direction in the bulk of the system can be chosen 
independently one site from the other, as by construction loops cannot be 
formed. The number of DSTs is  therefore 
$\mbox{$\cal{N}$}_{\mbox{\tiny $DST$}}\sim d^{N}$ 
(neglecting boundary terms). 
We have studied DSTs within our framework, selecting among all the  
single site and pair configurations for STs only those allowed by the 
directedness constraint (for instance, the number of single site 
configurations with $i$ edges is 
$m_s(i)=d \left(\begin{array}{c} d \\ i-1 \end{array} \right) $ for DSTs, as 
compared to 
$m_s(i)= i \left(\begin{array}{c} 2 d \\ i \end{array} \right) $ for STs). 
Similarly to the ST case, we have then  written a variational free energy from 
(\ref{free_en}). Minimising it numerically (subject to proper
constraints), we have indeed verified that the exact result is recovered, 
i.e. $ \mu_{\mbox{\tiny $DST$}}^{\mbox{\tiny $(B)$}}=d$.

In order to test the consistency of the scheme, we have studied the
case of spanning trees with energies depending on the number of
branchings at each site. To this end we have assigned a reduced energy
penalty $E_i$ to each site with $i$ branches. In the limit $E_1
\rightarrow \infty$ (and $E_i=0$ for $i>1$) tips are not allowed, so
that one should recover the case of space filling self-avoiding walk,
i.e.  Hamiltonian walks (HWs). We have verified that in fact when $E_1
\rightarrow \infty$ the only configuration which have a non zero
weight are those of a linear polymer. The resulting entropy moreover
coincides with that of HWs in the Bethe approximation \cite{lise}.  
Our Bethe approximation suggests a smooth crossover from STs to HWs,
with no sign of discontinuity in the free energy.
We have also verified, in the case $q=4$, that the HW limit can be
reached by sending simultaneously $E_3$ and $E_4$ to infinity, thus
forbidding the occurrence of branchings.

We now turn to the problem of LT collapse. 
Just as for linear polymers, the collapse transition 
is driven by an attractive interaction  $\beta$  between  nearest neighbour 
contacts\footnote{A contact  is defined as a pair of nearest neighbour 
vertices of the tree which are not linked by an edge}. 
The partition function of the so called {\em t-model} reads
\begin{equation}
\label{part_lt}
Z_{\mbox{\tiny $N$}} = 
\sum _{c \ge 0} t_{\mbox{\tiny $N$}}(c) e^{\beta c}
\end{equation}
where $t_{\mbox{\tiny $N$}}(c)$ denotes the number of trees with $N$ sites 
and $c$ contacts.
Introducing a monomer fugacity $z$, the grand
canonical partition function reads therefore
\begin{equation}
\label{granpart}
\mbox{$\cal{Z}$}=\sum _{N=1}^{\infty} \sum _{c \ge 0} 
               z^N  t_{\mbox{\tiny $N$}}(c) e^{\beta c} 
\end{equation}
where the first sum is over all possible number of sites in the tree.
We then proceed analogously to the ST case and write a pair 
approximation for the free energy of the system as
\begin{eqnarray} 
\label{free_lt}
\tilde f_{\mbox{\tiny $LT$}}^{\mbox{\tiny $(B)$}} 
= & & - \ln z \sum_{i=1}^{q} m_s(i) s_i + 
               \frac{q}{2} \beta \sum_{i,j=1}^{q-1} m_e(i,j) e_{ij}+ 
 \nonumber \\ & &
       \frac{q}{2}\left\{ \sum _{i,j=0}^{q-1} m_e(i,j) e_{ij} \ln e_{ij}+
         2 \sum_{i=0}^{q-1} \sum_{j=1}^{q-1} m_f(i,j) 
       \vec{f}_{ij} \ln \vec{f}_{ij} 
             \right\} -
        (q-1) \left\{ \sum_{i=0}^{q} m_s(i) s_i \ln s_i \right\}   
\end{eqnarray}
where the symbols $s_i$, $e_{ij}$, $\vec{f}_{ij}$ have similar meanings
as in eq.~(\ref{free_st1}) (note that contrary to eq.~(\ref{free_st1}) 
the possibility of an  empty site must be considered in 
eq.~(\ref{free_lt})).
The stable phase at given $\beta$ and $z$ is obtained by
minimising the 
free energy $\tilde f_{\mbox{\tiny $LT$}}^{\mbox{\tiny $(B)$}}$ subject to
normalisation and consistency conditions on the probability variables
(see previous discussion for STs).
In this case it is not possible to express the solution in a simple, 
analytical form and one has to fully resort to numerical methods, e.g. the
natural iteration method \cite{kik2}.

\begin{figure}[hb]
\epsfxsize=3.7in
\centerline{ \epsffile{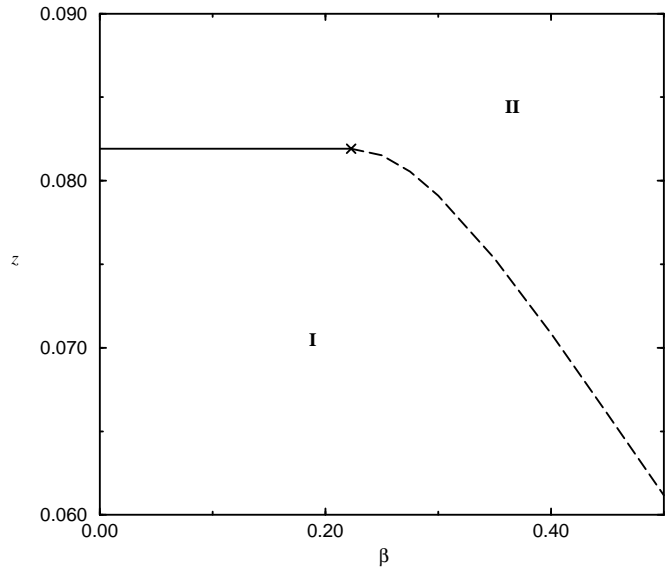}}
\caption[2]{\label{pd_3_fig}
Phase diagram of self-interacting lattice trees as a function of $\beta$ 
and $z$ for $d=3$. The average number of bonds in a tree is finite (infinite) 
in region I (II).
The continuous (dashed) line is a second (first)
order transition. The cross marks the tricritical point
($\beta _{\protect\mbox{\protect\tiny $\theta$}}  \approx 0.224$ and
$z _{\protect\mbox{\protect\tiny $\theta$}} \approx 0.08192$).}
\end{figure}

We report the complete phase diagram for $d=3$ in fig.~\ref{pd_3_fig},
as a function of $z$ and $\beta$. There are two distinct phases:
a zero density phase, where the average number of edges in a tree is finite, 
and a finite  density phase, where instead this number is infinite. 
These two phases are separated by a transition line $z_c(\beta)$ which 
could either be  first order (corresponding to a finite jump in density) 
or second order (corresponding to a continuous change in density). 
The tricritical point where the two lines merge is the 
$\theta$ point, $\beta _{\theta}$, i.e. the point where the collapse  
transition of the LT occurs. The structure of the infinite LT along the
line $z_c(\beta)$ is expanded for $\beta < \beta_{\theta}$ and 
compact for $\beta > \beta_{\theta}$.
We obtain $\beta_{\theta} \approx 0.406$ in $d=2$ and  
$\beta_{\theta}\approx 0.224$ in $d=3$. 
Recent estimates from extensive Monte Carlo simulations on the collapse of
lattice trees\cite{madras} yield the values
$\beta _{\theta}=0.699 \pm 0.052$
and $\beta _{\theta}=0.346 \pm 0.017$, respectively in $d=2$ and 
$d=3$. 

The phase diagram for LTs appears to be identical to the one for SAWs 
(see e.g. fig.~2 in \cite{lise}), except for a rescaling of the fugacity $z$ 
by a factor $\displaystyle{\left(\frac{q-1}{q-2}\right)^{q-2}}$. In 
particular the numerical values $\beta _{\theta}$ of the collapse transition 
of the polymers coincide  and, in both cases, $z_c$ does not depend on 
$\beta$, as long as $\beta < \beta_{\theta}$. This similarity is an 
intriguing result for which at the moment we don't have any plausible 
explanation. 
We have also investigated a generalisation of the partition function 
(\ref{part_lt}) by including an energy term $E_1$ which penalises  
configurations with tips. At any given $E_1$ the phase diagram in $\beta$ 
and $z$ shares 
similar features with the phase diagram  for LTs or SAWs. 
The difference is just restricted to a scale factor in $z$ which is determined 
by the value of $E_1$. This means in particular that the
value of $\beta _{\theta}$ does not depend on $E_1$, at least at this 
level of the approximation.
Similarly to the ST case, we have also verified that in the limit 
$E_1  \rightarrow \infty$, where tips are not allowed, the phase diagram of 
a linear polymer is recovered.

It is interesting to remark that in the special case $\beta=0$ we numerically
obtain for the connective constant of a LT 
$\mu_{\mbox{\tiny $LT$}}=z_c(0)^{-1}$ a result which agrees with the 
analytical expression
\begin{equation}
\mu_{\mbox{\tiny $LT$}}^{(\mbox{\tiny $B$})}=
\frac{(q-1)^{(q-1)}}{(q-2)^{(q-2)}}
\end{equation}
This expression was indeed derived  in \cite{fisher} by studying LTs 
on a Bethe lattice with coordination number $q$ and specializes to 
$\mu_{\mbox{\tiny $LT$}}^{(\mbox{\tiny $B$})}=6.75$ for $q=4$ and 
$\mu_{\mbox{\tiny $LT$}}^{(\mbox{\tiny $B$})}=12.21$ for $q=6$.
The most accurate numerical estimates\cite{gaunt_1} for the connective constant
of  $LTs$ are $\mu_{\mbox{\tiny $LT$}}\approx 5.14$ for the square lattice 
and $\mu_{\mbox{\tiny $LT$}} \approx 10.50$ for the cubic lattice.

In summary, we have introduced a novel variational technique to investigate
the behaviour of branched polymers. It is based on the cluster variation 
method with nearest neighbour pairs as maximal clusters. It can therefore be 
viewed as the natural formulation of the Bethe approximation for these 
systems and it provides the starting point on which systematic improvements 
can be made by applying the CVM recipe \cite{kik1,an} to clusters larger than
the NN pair (the first step being the plaquette). Also, it should be possible 
to deal with different branched polymer representations, such as weakly and 
strongly embedded lattice animals.

The method yields exact results in the trivial case of directed
spanning trees and extremely accurate ones for the enumeration of
spanning trees, where the exact result is still available. Introducing
branching energies in the spanning tree problem we have shown that the
Hamiltonian walk limit is correctly recovered. Finally, we have
studied the collapse transition of branched polymers, obtaining
estimates for the $\theta$ point which compare reasonably well with the
most accurate simulations. 

We thank A. Maritan for useful discussions. SL acknowledges EPSRC for 
financial support through a  postdoctoral fellowship.

\vskip 0.2cm

\noindent
$^a$ Electronic address:       Paolo.DeLosRios@unifr.ch  \\
$^b$ Electronic address:       lises@ic.ac.uk \\
$^c$ Electronic address:       alex@athena.polito.it

\end{document}